\documentclass[12pt,a4paper,epsf]{article}
\usepackage{graphics}
\usepackage{amssymb,amsmath}
\usepackage[dvips]{lscape,graphicx}
\usepackage{cite}

\newcommand{\bc}{\begin{center}}
\newcommand{\ec}{\end{center}}
\newcommand{\bd}{\begin{displaymath}}
\newcommand{\ed}{\end{displaymath}}
\newcommand{\be}{\begin{equation}}
\newcommand{\ee}{\end{equation}}
\newcommand{\ba}{\begin{array}}
\newcommand{\ea}{\end{array}}
\newcommand{\bea}{\begin{eqnarray}}
\newcommand{\eea}{\end{eqnarray}}
\newcommand{\bt}{\begin{tabular}}
\newcommand{\et}{\end{tabular}}
\newcommand{\un}{\underline}
\newcommand{\ov}{\overline}
\newcommand{\bp}{\begin{picture}}
\newcommand{\ep}{\end{picture}}
\newcommand{\bfi}{\begin{figure}}
\newcommand{\efi}{\end{figure}}

\begin{document}

\hyphenation{ }

\title{\huge \bf {Dark Energy and Dark Matter, Mirror World and $\huge \bf E_6$ Unification}}
\author{\Large \bf
C.R.~Das ${}^{1}$ \footnote{\large\,
crdas@pku.edu.cn}
L.V.~Laperashvili ${}^{2}$ \footnote{\large\, laper@itep.ru} , \\[5mm]
\itshape{${}^{1}$ \large Center for High Energy Physics, Peking University, Beijing, China}\\[0mm]
\itshape{${}^{2}$ \large The Institute of Theoretical and
Experimental Physics, Moscow, Russia}}

\date{}

\maketitle

\thispagestyle{empty}

\vspace{1cm}

\bc { \large \bf A talk presented at the Conference\\
 of Russian
Academy of Sciences:\\[5mm]

{\Large \bf Fundamental Interactions Physics}\\[7mm]

ITEP, Moscow, Russia\\[3mm] Nov 26-30, 2007}

\vspace{2cm}

{\Large \bf Speaker - Larisa Laperashvili}

\ec

\clearpage\newpage

\begin{abstract}
In the present talk we have developed a concept of parallel
ordinary (O) and mirror (M) worlds. We have shown that in the case
of a broken mirror parity (MP), the evolutions of fine structure
constants in the O- and M-worlds are not identical. It is assumed
that $E_6$-unification inspired by superstring theory restores the
broken MP at the scale $\sim 10^{18}$ GeV, what unavoidably leads
to the different $E_6$-breakdowns at this scale: $E_6 \to
SO(10)\times U(1)_Z$ - in the O-world, and $E'_6 \to SU(6)'\times
SU(2)'_Z$ - in the M-world. Considering only asymptotically free
theories, we have presented the running of all the inverse gauge
constants $\alpha_i^{-1}$ in the one-loop approximation. Then a
'quintessence' scenario is discussed for the model of accelerating
universe. Such a scenario is related with an axion ('acceleron')
of a new gauge group $SU(2)'_Z$ which has a coupling constant
$g_Z$ extremely growing at the scale $\Lambda_Z\sim 10^{-3}$ eV.

\end{abstract}

\clearpage\newpage

\setcounter{page}{1}

\Large \bf

{\huge \bf \un{Contents:}}

\vspace{1cm}

\begin{itemize}

\item[{\bf 1.}] Introduction: Superstring theory and a mirror
world.

\item[{\bf 2.}] Particle content in the ordinary and mirror
worlds.

\item[{\bf 3.}] Gauge coupling constant evolutions in the ordinary
world.

~~~{\large \bf 3.1. Standard Model and Minimal Supersymmetric
Standard Model.}

~~~{\large \bf 3.2. Left-right symmetry, SO(10) and $\large \bf
E_6$-unification. }

\item[{\bf 4.}] Mirror world with broken mirror parity.

~~~{\large \bf 4.1. Gauge coupling constant evolutions in the
mirror SM and MSSM.}

~~~{\large \bf 4.2. Mirror gauge coupling constant evolutions from
$\large \bf SU(6)$ to the $\large \bf E_6$-unification.}

\item[{\bf 5.}] A new mirror gauge group $\Large \bf SU(2)'_Z$.

~~~{\large \bf 5.1. Particle content of the $\large \bf SU(2)'_Z$
gauge group.}

~~~{\large \bf 5.2. The axion potential.}

~~~{\large \bf 5.3. A new cosmological scale $\large \bf
\Lambda_Z\approx 3\times 10^{-3}$ eV  .}

\item[{\bf 6.}] The gauge group $\Large \bf SU(2)'_Z$ and the
'quintessence' model of our universe.

~~~{\large \bf 6.1. Dark energy and cosmological constant.}

~~~{\large \bf 6.2. Dark matter.}

\item[{\bf 7.}] Conclusions.

\end{itemize}

\clearpage\newpage

\pagenumbering{arabic}

\large \bf

\section{\un{Introduction: Superstring theory and a mirror world.} }

\vspace{1cm}

The present investigation is based on the following corner stones
of theory:\\

$\bullet$ Grand Unified Theories (GUTs) are inspired by the
ultimate theory of superstrings, which gives the possibility of
unifying all fundamental interactions including gravity:\\

{\Large \bf \it M.B.~Green, J.H.~Schwarz and E.~Witten,}
Superstring theory, Vol. 1,2, Cambridge University Press,
Cambridge, 1988.\\

$\bullet$ There exists a mirror world, which is parallel to our
ordinary world:\\

{\Large \bf \it T.D.~Lee and C.N.~Yang,} Phys.Rev. {\bf 104}, 254
(1956);\\

{\Large \bf \it I.Yu.~Kobzarev, L.B.~Okun and I.Ya.~Pomeranchuk,}
Yad.Fiz. {\bf 3}, 1154 (1966) [Sov.J.Nucl.Phys. {\bf 3}, 837
(1966)].\\

$\bullet$ The mirror parity MP is broken:\\

{\Large \bf "The only good parity ... is a broken parity!"}\\[3mm]

{\Large \bf \it Z.~Berezhiani, A.~Dolgov and R.N.~Mohapatra,}
Phys.Lett. B {\bf 375}, 26 (1996);

{\Large \bf \it Z.~Berezhiani,} Acta Phys.Pol. B {\bf 27}, 1503
(1996);

{\Large \bf \it Z.G.~Berezhiani and R.N.~Mohapatra,} Phys.Rev. D
{\bf 52}, 6607 (1997).

\clearpage\newpage

\section*{\un{Introduction: Superstring theory and a mirror world.}}

\vspace{1cm}

Superstring theory is a paramount candidate for the unification of
all fundamental gauge interactions with gravity.\\

Superstrings are free of gravitational and Yang-Mills\\ anomalies
if a gauge group of symmetry is $$\large \bf SO(32)\quad \rm{or}
\quad \large \bf E_8\times E_8.$$

The 'heterotic' superstring theory $\large \bf E_8\times E'_8$ was
suggested
as a more realistic model for unification:\\

\bc {\Large \bf \it D.J.~Gross, J.A.~Harvey, E.~Martinec and
R.~Rohm,}

 Phys.Rev.Lett. {\bf 54}, 502 (1985); Nucl.Phys. {\bf
B256},
253 (1985). \\[5mm]
{\Large \bf \it M.B.~Green, J.H.~Schwarz and E.~Witten,}

Superstring theory, Vol. 1,2, Cambridge University Press,
Cambridge, 1988.\\[7mm]
\ec This ten-dimensional Yang-Mills theory can undergo spontaneous
compactification:\\[5mm]
{\Large \bf \it The integration over 6 compactified dimensions of
the $\large \bf E_8$ superstring theory leads to the effective
theory with the $\large \bf E_6$-unification in four-dimensional
space.}

\clearpage\newpage

\section*{\un{Introduction: Superstring theory and a mirror world.}}

\vspace{1cm}

In the present investigation:

See:\\
{\Large \bf \it C.R.~Das and L.V.~Laperashvili,} Mirror World with
Broken Mirror Parity, $E_6$ Unification and
Cosmology,

to be published in Phys.Rev. D. \\[3mm]
we consider the old concept:\\[3mm]
{\Large \bf \it there exists in Nature a 'mirror' (M) world
(hidden sector) parallel to our ordinary (O) world.}\\[5mm]
This M-world is a mirror copy of the O-world and contains the same
particles and their interactions as our visible world.\\[5mm]
Observable elementary particles of our O-world have left-handed
(V-A) weak interactions which violate P-parity. If a hidden mirror
M-world exists, then mirror particles participate in the
right-handed (V+A) weak interactions and have an opposite
chirality.\\[5mm]
Lee and Yang were first who suggested such a duplication of the
worlds which restores the left-right symmetry of Nature:\\[5mm]
{\Large \bf \it T.D.~Lee and C.N.~Yang,} Phys.Rev. {\bf 104}, 254
(1956);\\[5mm]
The term 'Mirror World' was introduced by Kobzarev, Okun and
Pomeranchuk:\\[5mm]
{\Large \bf \it I.Yu.~Kobzarev, L.B.~Okun and I.Ya.~Pomeranchuk,}
Yad.Fiz. {\bf 3}, 1154 (1966) [Sov.J.Nucl.Phys. {\bf 3}, 837
(1966)].\\[5mm]
They have investigated a lot of phenomenological implications of
such parallel worlds.\\[5mm]
The idea of the existence of visible and mirror worlds became very
attractive in connection with a superstring theory described by
$\large \bf E_8\times E'_8$.

\clearpage\newpage

\section{\un{Particle content in the ordinary and mirror worlds.}}

We can describe the ordinary and mirror worlds by a minimal
symmetry $$\Large \bf G_{SM}\times G'_{SM}, \quad where$$
$$\Large \bf G_{SM} = SU(3)_C\times SU(2)_L\times U(1)_Y$$ stands
for the Standard Model (SM) of observable particles: three
generations of
quarks and leptons and the Higgs boson.\\
Then $$\Large \bf G'_{SM} = SU(3)'_C\times SU(2)'_L\times
U(1)'_Y$$ is its mirror gauge counterpart having three generations
of mirror quarks and leptons and the mirror Higgs boson.\\[5mm]
The M-particles are singlets of $\Large \bf G_{SM}$ and
O-particles
are singlets of $\Large \bf G'_{SM}$.\\[5mm]
{\Large \bf \it These different O- and M-worlds are coupled only
by gravity (or maybe other very weak interaction).}\\[5mm]
Including Higgs bosons $\Large \bf \phi$ we have the following SM
content of the O-world: $$\Large \bf \rm{\Large \bf L-set}: \Large
\bf \quad (u,d,e,\nu,\tilde u,\tilde d,\tilde e,\tilde
N)_L\,,\phi_u,\,\phi_d ;$$ $$\Large \bf {\rm \Large \bf \tilde
R-set}: \Large \bf \quad (\tilde u,\tilde d,\tilde e,\tilde
\nu,u,d,e,N)_R\,,\tilde \phi_u,\,\tilde \phi_d;$$ with
antiparticle fields: $\Large \bf \tilde \phi_ {u,d} =
\phi^*_{u,d},\,\,$ $\Large \bf \tilde \psi_R =
C\gamma_0\psi_L^*\,\,$ and $\Large \bf \tilde \psi_L =
C\gamma_0\psi_R^*.$

Considering the minimal symmetry $\Large \bf G_{SM}\times G'_{SM}$
we have the following particle content in the M-sector:
$$ \Large \bf \rm{\Large \bf L'-set}: \Large \bf \quad \Large \bf
(u',d',e',\nu',\tilde u',\tilde d',\tilde e',\tilde
N')_L\,,\phi'_u,\,\phi'_d ;$$ $$\Large \bf {\rm \Large \bf
 \tilde R'-set}: \Large \bf \quad (\tilde u',\tilde d',\tilde e',\tilde
\nu',u',d',e',N')_R\,,\tilde \phi'_u,\,\tilde \phi'_d.$$ In
general, we can consider a supersymmetric theory when\\ $\Large
\bf G\times G'$ contains grand unification groups: $\Large \bf
SU(5)\times SU(5)'$,\\ $\Large \bf SO(10)\times SO(10)',\,\,$
$\Large \bf E_6\times E_6'$, etc.

\clearpage\newpage

\section{\un{Gauge coupling constant evolutions in the
O-world.}}

In the present paper we consider the running of all the gauge
coupling constants in the SM and its extensions which is well
described by the one-loop approximation of the renormalization
group equations (RGEs) from the Electroweak (EW) scale up to the
Planck scale.\\[3mm]
For energy scale $ \Large \bf \mu \ge M_{ren}$, where $ \Large \bf
 M_{ren}$ is the renormalization scale, we have the following
evolution for the inverse fine structure constants $ \Large \bf
 \alpha_i^{-1}$ given by RGE in the one-loop approximation: $$
 \Large\bf
 \alpha_i^{-1}(\mu) = \alpha_i^{-1}(M_{ren}) + \frac{b_i}{2\pi}t,$$ where
 $$ \Large \bf \alpha_i =\frac {g^2_i}{4\pi}, $$ $ \Large \bf g_i$ are
gauge coupling constants and $$ \Large \bf
t=ln\left(\frac{\mu}{M_{ren}}\right).$$

We have assumed that the following chain of symmetry groups exists
in the ordinary world:
$$\Large \bf  SU(3)_C\times
SU(2)_L\times U(1)_Y \to  [SU(3)_C\times SU(2)_L\times
U(1)_Y]^{SUSY}
$$ $$\to \Large \bf
 SU(3)_C\times SU(2)_L \times SU(2)_R\times U(1)_X\times
 U(1)_Z$$
 $$ \Large\bf \to SU(4)_C\times SU(2)_L \times SU(2)_R\times U(1)_Z $$ $$\to
\Large \bf SO(10)\times U(1)_Z \to E_6.$$

\clearpage\newpage

\subsection{\un{Standard Model and Minimal Supersymmetric Standard
Model.}}

\vspace{1cm}

We start with the SM in our ordinary world.

In the SM for energy scale $\Large \bf \mu \ge M_t$ (here $\Large
\bf M_t$ is the top quark pole mass) we have the following
evolutions (RGEs) for the inverse fine structure constants $\Large
\bf \alpha_i^{-1}$ ($\Large \bf i=1,2,3$ correspond to the $\Large
\bf U(1),\,\,SU(2)_L$ and $\Large \bf SU(3)_C$ groups of the SM):\\[3mm]
{\Large \bf \it C.~Ford, D.R.T.~Jones, P.W.~Stephenson, M.B.~Einhorn,}\\
Nucl.Phys. B {\bf 395}, 17 (1993),\\[5mm]
which are revised using updated experimental results:\\[5mm]
{\Large \bf \it C.R.~Das, C.D.~Froggatt, L.V.~Laperashvili and
H.B.~Nielsen,}\\ Mod.Phys.Lett. A {\bf 21}, 1151 (2006)\\[5mm]
{ \Large \bf \it C.D.~Froggatt, L.V.~Laperashvili, H.B.~Nielsen,}
Phys.Atom.Nucl. {\bf 69}, 67 (2006); Yad.Fiz. {\bf 69}, 3 (2006).\\[5mm]

$$\Large \bf
      \alpha_1^{-1}(t) = 58.65 \pm 0.02 - \frac{41}{20\pi}t,
$$ $$ \Large \bf      \alpha_2^{-1}(t) = 29.95 \pm 0.02 +
\frac{19}{12\pi}t,$$
$$ \Large \bf      \alpha_3^{-1}(t) = 9.17 \pm 0.20 +
\frac{7}{2\pi}t,
$$\\
where  $$\Large \bf t=ln\left(\frac{\mu}{M_t}\right).$$ We have
used the central value of the top quark mass: $$ \Large \bf
M_t\approx 174\,\, GeV.$$\\

\subsection*{\un{Standard Model and Minimal Supersymmetric Standard
Model.}}

\vspace{1cm}

The Minimal Supersymmetric Standard Model (MSSM) \bc (which
extends the conventional SM) \ec gives the evolutions for $\Large
\bf \alpha_i^{-1}$\\

($\Large \bf i=1,2,3\,\,$ for $\Large \bf U(1),\,SU(2),\,SU(3)$
groups)\\

from the supersymmetric scale
$\Large \bf M_{SUSY}$ up to the seesaw scale $\Large \bf M_R$.\\

Figs.~1,3 present by red lines the SM and MSSM evolutions, which
are given by the following MSSM slopes: $$\Large \bf
           b_1 = - \frac{33}{5} = - 6.6, \qquad
     b_2 = -1, \qquad b_3 = 3.$$\\
These evolutions are shown from $\Large \bf M_t$ up to the scale
$\Large \bf M_{SUSY}$, where $$\Large \bf x = log_{10}\mu
\,(\rm{\Large \bf GeV}),\quad \Large \bf t = x\cdot ln10 - ln
M_t.$$

In Figs.~1-4 we have presented examples with the following
scales:

Fig.~1,2 -- $\Large \bf 10 \,\,\rm{\Large \bf TeV},$

Fig.~3,4 --  $\Large \bf 1 \,\,\rm{\Large \bf TeV},$\\

 and
$$\Large \bf M_R \sim 10^{14}\,\,{\rm{or}}\,\,10^{15}\,\, \rm{\Large \bf GeV}.$$\\ Here and below red lines
correspond to the ordinary world.

 \clearpage\newpage

\subsection{\un{Left-right symmetry, SO(10) and $\large \bf
E_6$-unification.}}

\vspace{1cm}

At the seesaw scale $\Large \bf M_R$ the heavy right-handed
neutrinos appear,
 and the following supersymmetric left-right
symmetry originates:
 $$\Large \bf
 SU(3)_C\times SU(2)_L\times
SU(2)_R\times U(1)_X\times U(1)_Z. $$

Considering the running of coupling constants we have the
following slopes:
$$\Large \bf
       b_X = b_1 = - 6.6, \quad b_Z = - 9, \quad b_3 = 3.
$$
Also  the running for $\Large \bf SU(2)_L\times SU(2)_R$ is given
by the slope: $$\Large \bf
                 b_{22} = - 2.$$
Then we have the following evolution: $$\Large \bf
 \alpha_{22}^{-1}(\mu) = \alpha_{22}^{-1}(M_R) +
 \frac{1}{\pi}\ln\frac{\mu}{M_R},$$
with the following relation: $$\Large \bf
         \alpha_{22}^{-1}(M_R) = \alpha_2^{-1}(M_R).$$
The next step is an assumption that the group by Pati and Salam
originates at the scale $\Large \bf M_4$ giving the following
extension of the group: $$\Large \bf
 SU(3)_C\times SU(2)_L\times
SU(2)_R\times U(1)_X\times U(1)_Z $$ $$\Large \bf \to
SU(4)_C\times SU(2)_L\times SU(2)_R\times U(1)_Z. $$

{\Large \bf \it J.~Pati and A.~Salam,} Phys.Rev. D {\bf 10}, 275
(1974).\\[3mm]
The scale $\Large \bf M_4$ is given by the intersection of $\Large
\bf SU(3)_C$ with $\Large \bf U(1)_X$: $$\Large \bf
        \alpha_3^{-1}(M_4) = \alpha_X^{-1}(M_4).   $$

\clearpage\newpage

\subsection*{\un{Left-right symmetry, SO(10) and $\large \bf
E_6$-unification.}}

\vspace{1cm}

Considering only the minimal content of the scalar Higgs fields,
we obtain the following slope for the running of $\Large \bf
\alpha_4^{-1}(\mu)$: $$\Large \bf
       b_4 =  5. $$
The intersection of $\Large \bf \alpha_4^{-1}(\mu)$ with the
running of $\Large \bf \alpha_{22}^{-1}(\mu)$ leads to the scale
$\Large \bf M_{GUT}$ of the SO(10)-unification:$$\Large \bf
        SU(4)_C\times SU(2)_L \times SU(2)_R \to SO(10),$$
$$\Large \bf
 \alpha_4^{-1}(M_{GUT}) = \alpha_{22}^{-1}(M_{GUT}). $$
Then we deal with the running of the SO(10) inverse gauge constant
$\Large \bf \alpha_{10}^{-1}(\mu)$, which runs from the scale $
\Large \bf M_{GUT}$ up to the scale $\Large \bf M_{SGUT}$ of the
super-unification $\Large \bf E_6$: $$\Large \bf
     SO(10)\times U(1)_Z \to E_6. $$
The slope of this running is: $$
   \Large \bf      b_{10} = 1.$$
Then we have the following running:
$$ \Large \bf
 \alpha_{10}^{-1}(\mu)
 = \alpha_{10}^{-1}(M_{GUT}) +
 \frac{1}{2\pi}\ln \frac{\mu}{M_{GUT}},$$
which is valid up to the $ \Large \bf  M_{SGUT}= M_{E6}\sim
10^{18}$ GeV.\\[3mm]

All evolutions of the corresponding fine structure constants are
given in Figs.~1-4: \bc (O-world -- red lines;\,\, M-world -- blue
lines).\ec

Here Figs.~2 and 4 show the running of gauge coupling constants
near the scale of the $\Large \bf E_6$-unification (for $\Large
\bf x\ge 15$).

\clearpage\newpage

\section{\un{Mirror world with broken mirror parity.}}

\vspace{1cm}

In general case the mirror parity MP is not conserved,\\
and the ordinary and mirror worlds are not identical:\\

{\Large \bf \it Z.~Berezhiani, A.~Dolgov and R.N.~Mohapatra,}
Phys.Lett. B {\bf 375}, 26 (1996);

{\Large \bf \it Z.~Berezhiani,} Acta Phys.Pol. B {\bf 27}, 1503
(1996);

{\Large \bf \it Z.G.~Berezhiani and R.N.~Mohapatra,} Phys.Rev. D
{\bf 52}, 6607 (1997).\\
If O- and M-sectors are described by the minimal symmetry group
$$ \Large \bf       G_{SM}\times G'_{SM} $$
with the Higgs doublets $\phi$ and $\phi'$, respectively,\\
then in the case of non-conserved MP the VEVs of $\phi$ and
$\phi'$ are
not identical: $\Large \bf v\neq v'$.\\

Following Berezhiani-Dolgov-Mohapatra,  we assume that $$\Large
\bf v'>>v$$ and introduce the parameter characterizing the
violation of MP:
$$\Large \bf
        \zeta = \frac{v'}{v} >> 1. $$
Then the masses of fermions and massive bosons in the mirror world
are scaled up by the factor $\Large \bf \zeta$:
$$\Large \bf
               m'_{q',l'} = \zeta m_{q,l},  $$
$$\Large \bf
                 M'_{W',Z',\phi'} = \zeta M_{W,Z,\phi}, $$
but photons and gluons remain massless in both worlds.

\clearpage\newpage

\section*{\un{Mirror world with broken mirror parity.}}

\vspace{1cm}

Let us consider now the following expressions:
$$\Large \bf
    \alpha_i^{-1}(\mu) = \frac{b_i}{2\pi}\ln
    \frac{\mu}{\Lambda_i}$$
--- in the O-world, and
$$\Large \bf {\alpha'}_i^{-1}(\mu) = \frac{b'_i}{2\pi}\ln
    \frac{\mu}{\Lambda'_i}$$

--- in the M-world.\\

A big difference between the Electroweak scales $\Large \bf v$ and
$\Large \bf v'$ will not cause a big difference between scales
$\Large \bf \Lambda_i$ and $\Large \bf \Lambda'_i$: $$ \Lambda'_i
= \xi \Lambda_i   \quad \rm{with}\quad \xi > 1. $$ The values of
$\,\,\zeta \,\,$ and $\xi$ were estimated by astrophysical
implications \bc (by Berezhiani-Dolgov-Mohapatra), \ec which gave:
$$ \Large \bf             \zeta\approx 30 \quad \rm{and} \quad
                         \xi\approx 1.5. $$
As for the neutrino masses, the same authors have shown that the
theory with broken mirror parity leads to the following relations:
$$\Large \bf
      m'_{\nu} = \zeta^2 m_{\nu}, $$
$$\Large \bf
      M'_{\nu} = \zeta^2 M_{\nu}, $$
where $\Large \bf m_{\nu}$ are light left-handed and $\Large \bf
 M_{\nu}$ are heavy right-handed neutrino masses in the O-world,
and $\Large \bf m'_{\nu},M'_{\nu}$ are the corresponding neutrino
masses in the M-world.\\
The last relation gives the following relation for seesaw scales:
$$\Large \bf
             M'_R = \zeta^2 M_R.$$

 \clearpage\newpage

\subsection{\un{Gauge coupling constant evolutions in the mirror SM and
MSSM.}}

\vspace{1cm}

In the SM of the M-sector we have the following evolutions:
$$\Large \bf
 {(\alpha')}_i^{-1}(\mu) = {(\alpha')}_i^{-1}(M'_t) + \frac{b_i}{2\pi}t'
 = \frac{b_i}{2\pi}\ln \frac{\mu}{\Lambda'_i},
$$ where $$\Large \bf
      {(\alpha')}_i^{-1}(M_t) = \alpha_i^{-1}(M_t) - \frac{b_i}{2\pi}\ln \xi,
$$or $$\Large \bf
           {(\alpha')}_i^{-1}(M'_t) = \alpha_i^{-1}(M_t).
$$
In the M-world the scales $\Large \bf \Lambda'_i$ are different
with $\Large \bf \Lambda_i$, but O- and M-slopes are
identical:$$\Large \bf
         b'_i = b_i.$$
Finally, we obtain the following SM running of gauge coupling
constants in the mirror world:

 1)$$\Large \bf
    {(\alpha')}_1^{-1}(\mu) = 58.65 \pm 0.02
                      - \frac{41}{20\pi}t',
$$

2) $$\Large \bf
    {(\alpha')}_2^{-1}(\mu) = 29.95 \pm 0.02 +
    \frac{19}{12\pi}t',
$$

3) $$\Large \bf
    {(\alpha')}_3^{-1}(\mu) = 9.17 \pm 0.20 + \frac{7}{2\pi}t',$$
where $$\Large \bf t' = ln{\left(\frac{\mu}{M'_t}\right)}.$$

The pole mass of the mirror top quark is $$\Large \bf M'_t =\zeta
M_t.$$

\clearpage\newpage

\subsection*{\un{Gauge coupling constant evolutions in the mirror SM and
MSSM.}}

\vspace{1cm}

If the Minimal Supersymmetric Standard Model (MSSM) extends the
mirror SM , then mirror sparticle masses obey the following
relation:
$$\Large \bf
         \widetilde {m'} = \zeta \widetilde {m},  $$
and the mirror SUSY-breaking scale is larger:
$$\Large \bf
      M'_{SUSY} = \zeta M_{SUSY}.  $$
The mirror MSSM gives the evolutions for $\Large \bf
{\alpha'}_i^{-1}(\mu)$ ($i=1,2,3$) from the supersymmetric scale
$\Large \bf M'_{SUSY}$  up to the mirror GUT scale $\Large \bf
M'_{GUT}$.\\[3mm]

A seesaw scale $\Large \bf M'_R$ in the M-world is given in the
previous Subsection. For $\Large \bf \zeta = 3$:
        $$\Large \bf  M'_R = \zeta^2 M_R\approx 10^3 M_R. $$\\
Now if $\Large \bf M_R \sim 10^{14}$ GeV, then $\Large \bf M'_R
\sim 10^{17}$ GeV, and a seesaw scale is close to the superGUT
scale of the $E_6$-unification.\\[3mm]
This means that mirror heavy right-handed neutrinos appear at the
scale $\Large \bf \sim 10^{17}$ GeV.

Figs.~1-4 present by blue lines the mirror MSSM evolutions  of
$\Large \bf {\alpha'}_i^{-1}(\mu)$ ($\Large \bf i=1,2,3$).\\
In Figs.~1-4 we have presented (by blue lines) examples of the
mirror MSSM evolutions with the scales $$\Large \bf M'_{SUSY} =
10\,\, \rm{\Large \bf TeV} \,\, \rm{\Large \bf and}\,\,\Large \bf
300 \,\, \rm{\Large \bf TeV},$$ and $\Large \bf M'_R\sim 10^{17}$
GeV;\\ $\Large \bf \zeta = 10$ -- for $\Large \bf M_{SUSY} = 1$
TeV, and $\Large \bf \zeta = 30$ -- for $\Large \bf M_{SUSY} = 10$
TeV.

\clearpage\newpage

\subsection{\un{Mirror gauge coupling constant evolutions}}{\large \bf \un{ from $\large
\bf SU(6)$ to the $\large \bf E_6$-unification.}}

\vspace{1.5cm}

Let us consider now the extension of the MSSM in the mirror world.

The first step of such an extension is:
$$\Large \bf  [SU(3)'_C\times
SU(2)'_L\times U(1)'_Y]_{MSSM}$$ $$ \to \Large \bf
[SU(3)'_C\times SU(2)'_L\times U(1)'_X\times U(1)'_Z]_{MSSM},
$$\\
and then
$$\Large \bf  [SU(3)'_C\times
SU(2)'_L\times U(1)'_X]_{MSSM}$$ $$\Large \bf \to  SU(4)'_C\times
SU(2)'_L.$$\\

Assuming that the supersymmetric group $\Large \bf SU(4)'_C\times
SU(2)'_L$ originates at the scale $\Large \bf M'_4$, we find the
intersection of $\Large \bf SU(3)'_C$ with $\Large \bf U(1)'_X$:
$$\Large \bf
        {\alpha'}_3^{-1}(M'_4) = \alpha_X^{-1}(M'_4). $$\\
The gauge symmetry group $\Large \bf SU(4)'_C$ starts from the
scale $\Large \bf M'_4$ and runs up to the intersection with the
evolution $\Large \bf {(\alpha')}_2^{-1}(\mu)$ corresponding to
the supersymmetric group $\Large \bf SU(2)'_L$.\\

Here we have: $$\Large \bf
b_2=-1.$$\\
The point of this intersection is the scale $\Large \bf M'_{GUT}$.
\clearpage\newpage

\subsection*{\un{Mirror gauge coupling constant evolutions}}{\large \bf \un{ from $\large
\bf SU(6)$ to the $\large \bf E_6$-unification.}}

\vspace{1.5cm}

The scale $\Large \bf M'_{GUT}$ is given by the following
relation: $$\Large \bf
 {(\alpha')}_4^{-1}(M'_{GUT}) = {(\alpha')}_2^{-1}(M'_{GUT}). $$

At the mirror GUTscale $\Large \bf M'_{GUT}$ we obtain the $\Large
\bf SU(6)'$-unification if $\Large \bf U(1)'_Z$ also meets $\Large
\bf  SU(4)'_C$ and $\Large \bf SU(2)'_L$ at the same scale:
$$ \Large \bf
        SU(4)'_C\times SU(2)'_L\times U(1)'_Z  \to SU(6)'. $$
Here again $$\Large \bf b_Z=-9.$$

Then we consider the running of $\Large
\bf{(\alpha')}_6^{-1}(\mu)$ up to the superGUT scale $\Large \bf
M'_{SGUT} = M'_{E6}$: $$ \Large \bf {(\alpha')}_6^{-1}(\mu) =
{(\alpha')}_6^{-1}(M'_{GUT}) +
 \frac{11}{2\pi}\ln \frac{\mu}{M'_{GUT}},$$
where we have used the result$$\Large \bf
    b_6 = 11. $$
Calculating the slope $\Large \bf b_6$, we assumed the existence
of only minimal number of the Higgs fields, namely $\Large \bf h +
\bar h$, belonging to the fundamental representation \un 6 of the
$\Large \bf SU(6)'$ group.

Now it is obvious that we must find some unknown in the O-world
symmetry group $\Large \bf SU(2)'_Z$, which must help us to get
the desirable $\Large \bf E'_6$-unification in the M-world at the
superGUT scale $\Large \bf M'_{SGUT}$: $$ \Large \bf SU(6)'\times
SU(2)'_Z \to E'_6. $$

\clearpage\newpage

\subsection*{\un{Mirror gauge coupling constant evolutions}}{\large \bf \un{ from $\large
\bf SU(6)$ to the $\large \bf E_6$-unification.}}

\vspace{1.5cm}

In the present investigation we assume that at the very small
distances the mirror parity is restored and super-unifications
$\Large \bf E_6$ and $\Large \bf E'_6$, inspired by superstring
theory, are identical having the same $\Large \bf M_{SGUT}$:
$$\Large \bf M'_{SGUT} = M_{SGUT} = M_{E6}\sim
10^{18}\,\,GeV.$$
 By this reason, the
superGUT scale $\Large \bf M_{SGUT}$ may be fixed by the
intersection of the evolutions of gauge coupling constants in both
-- mirror and ordinary -- worlds, which from the beginning were
not identical.

 The scale $\Large \bf M_{SGUT}$ of the $\Large \bf E_6\times
E'_6$-unification is given by the following intersection: $$\Large
\bf
    \alpha_{10}^{-1}(M_{SGUT}) =  {(\alpha')}_6^{-1}(M_{SGUT}). $$
Finally, one can envision the following symmetry breaking chain in
the M-world:
$$\Large \bf
E'_6 \to SU(6)'\times SU(2)'_Z $$ $$\to SU(4)'_C\times
SU(2)'_L\times SU(2)'_Z\times U(1)'_Z $$
$$\Large \bf \to SU(3)'_C\times
SU(2)'_L\times SU(2)'_Z\times U(1)'_X \times U(1)'_Z $$
$$\Large \bf \to \left[SU(3)'_C\times
SU(2)'_L\times U(1)'_Y\right]\times SU(2)'_Z.
$$\\
Now it is quite necessary to understand if there  exists  the
group $\Large \bf SU(2)'_Z$ in the mirror world.\\[3mm]
{\Large \bf What it could be?}

\clearpage\newpage

\section{\un{A new mirror gauge group $\Large \bf SU(2)'_Z$.}}

\vspace{0.5cm}

The reason of our choice of the gauge group $\Large \bf  SU(2)'_Z$:\\[3mm]
See:
 {\Large \bf \it C.R.~Das and L.V.~Laperashvili,}
Mirror World with Broken Mirror Parity, $E_6$ Unification and
Cosmology,
submitted to Phys.Rev. D; ArXiv:\\[3mm]
was to obtain the correct running of $\Large \bf
{(\alpha')}_{2Z}^{-1}(\mu)$, which:\\[3mm]
{\Large \bf $\bullet$ leads to the new scale $\Large \bf
\Lambda_Z\sim 10^{-3}$ eV at
extremely low energies;}\\[3mm]
See: {\Large \bf \it H.~Goldberg,} Phys.Lett. B {\bf 492}, 153
(2000).\\ {\Large \bf \it P.Q.~Hung,} Nucl.Phys. B {\bf 747}, 55
(2006);
 J.Phys. A {\bf 40}, 6871 (2007);  arXiv: hep-ph/0707.2791.\\
{\Large \bf \it P.Q.~Hung and P.~Mosconi,} ArXiv: hep-ph/0611001.\\[3mm]
{\Large \bf $\bullet$ is consistent with the running of all
inverse gauge coupling constants in the
O- and M-worlds with broken mirror parity, considered in this investigation.}\\[3mm]
Only the following slopes are consistent with our aims:
$$\Large \bf
        b_{2Z} =\frac{13}{3}\approx 4.33\quad \rm{\Large \bf and}\quad \Large \bf b_{2Z}^{SUSY}=
        0.$$

\subsection{\un{Particle content of the $\large \bf SU(2)'_Z$ gauge
group.}}

The particle content of $\Large \bf  SU(2)'_Z$ is as follows:

\begin{itemize}

\item[{\bf 1.}] two doublets of fermions $\psi^{(Z)}_i$ and two
doublets of the 'messenger' scalar fields $\Large \bf
\phi^{(Z)}_i$ with $\Large \bf i = 1,2$, or

\item[{\bf 2.}] one triplet of fermions $\Large \bf  \psi^{(Z)}_f$
with $\Large \bf  f=1,2,3$, which are singlets under the SM, and
two doublets of the 'messenger' scalar fields $\Large \bf
\phi^{(Z)}_i$ with $\Large \bf i = 1,2$.

\item[{\bf 3.}] We also consider a complex singlet scalar field:
 $ \Large \bf  \varphi_Z = (1,1,0,1) $
under the symmetry group $$\Large \bf  G' =\left[ SU(3)'_C\times
SU(2)'_L\times U(1)'_Y\right]\times SU(2)'_Z.$$

\end{itemize}

\clearpage\newpage

\subsection*{\un{Particle content of the $\large \bf SU(2)'_Z$ gauge
group.}}

\vspace{0.5cm}

The so called 'messenger' fields $\Large \bf  \phi^{(Z)}$ carry
quantum numbers of both the SM' and $\Large \bf  SU(2)'_Z$ groups.
They have Yukawa couplings with SM' leptons and fermions $\Large
\bf  \psi^{(Z)}$.

All the SM' and SM particles are assumed to be singlets under
$\Large \bf  SU(2)'_Z$. Then we obtain the following evolutions:

1. for the region $\Large \bf M'_t \le \mu \le M'_{SUSY}$:
 $$\Large \bf
 {\alpha'}_{2Z}^{-1}(\mu) = {\alpha'}_{2Z}^{-1}(M'_t) +
 \frac{b_{2Z}}{2\pi}\ln\frac{\mu}{M'_t}\approx
 \frac{b_{2Z}}{2\pi}\ln\frac{\mu}{\Lambda_Z},
             $$

2. and for the region $\Large \bf M'_{SUSY}\le \mu \le M'_{SGUT}$:
$$ \Large \bf   {\alpha'}_{2Z}^{-1}(\mu) =
{\alpha'}_{2Z}^{-1}(M'_{SUSY}) +
 \frac{b_{2Z}^{SUSY}}{2\pi}\ln\frac{\mu}{M'_{SUSY}}.
   $$
Also we have the following relation: $$ \Large \bf
{\alpha'}_{2Z}^{-1}(M'_{SGUT}=M_{SGUT}) = \alpha^{-1}_{E6}.
                                  $$
In Figs.~1-4 we have shown the evolution $\Large \bf
 {\alpha'}_{2Z}^{-1}(\mu)$ given by blue lines for $$\Large \bf b_{2Z}=\frac{13}{3}$$ and
 $$\Large \bf b_{2Z}^{SUSY} = 0.$$

The total picture of the evolutions in the O- and M-worlds
is presented simultaneously in Figs.~1--4 for the cases:\\[3mm]
$ \Large \bf M_{SUSY} = 1\,\,\rm {\Large \bf and}\,\, \Large \bf
10$ TeV, $\Large \bf M_R\sim 10^{14}, 10^{15}$ GeV, $\Large \bf
M'_R\sim 10^{17}$ GeV, $\Large \bf \zeta = 10\,\, and\,\,
\zeta =30$.\\
It is obvious that respectively $\Large \bf
M'_{SUSY} = 10\,\,\rm {\Large \bf and}\,\, \Large \bf 300$ TeV.\\
Here $\Large \bf M_{SGUT}\approx 7\cdot 10^{17}$ GeV and $\Large
\bf \alpha_{SGUT}^{-1}\approx 27.64$\\ -- for Figs.~1,2 ($\Large \bf M_{SUSY}=10$ TeV),\\
$\Large \bf M_{SGUT}\approx 2.4\cdot 10^{17}$ GeV and $\Large \bf
\alpha_{SGUT}^{-1}\approx 26.06$\\
-- for Figs.~3,4 ($\Large \bf M_{SUSY}=1$ TeV).\\[3mm]
 Red lines correspond to the ordinary world,
 blue lines -- to the mirror world.\\[3mm]

\clearpage\newpage

\subsection{\un{The axion potential.}}

\vspace{1cm}

The Lagrangian corresponding to the group of symmetry
$$\Large \bf  G' = SU(3)'_C\times
SU(2)'_L\times SU(2)'_Z\times U(1)'_Y$$ exhibits a $\Large \bf
U(1)_A^{(Z)}$ global symmetry.\\[3mm]
The singlet complex scalar field $\Large \bf \varphi_Z$ was
introduced in theory with aim to reproduce the model of Peccei-Quinn (PQ) (well-known in QCD):\\[3mm]
{\huge \bf \it R.~Peccei and H.~Quinn,} Phys.Rev.Lett. {\bf 38},
1440 (1977).\\[3mm]
Then the potential: $$\Large \bf
     V = \frac{\lambda}{4}(\varphi^+_Z\varphi_Z - v_Z^2)^2
                    $$
gives the VEV for $\Large \bf  \varphi_Z$: $$ \Large \bf
<\varphi_Z> = v_Z.
$$
Representing the field $\Large \bf \varphi_Z$ as follows:
 $$ \Large \bf
      \varphi_Z = v_Z\exp(ia_Z/v_Z) + \sigma_Z,
$$
we obtain the following VEVs: $$ \Large \bf <a_Z> = <\sigma_Z> =
0.$$ A boson $\Large \bf a_Z$ (the imaginary part of a singlet
scalar field $\Large \varphi_Z$) is an axion, and could be a
massless Nambu-Goldstone (NG) boson if the $U(1)_A^{(Z)}$ symmetry
is not spontaneously broken.

However, the spontaneous breakdown of the global $\Large \bf
U(1)_A^{(Z)}$ by $SU(2)'_Z$ instantons gives masses to fermions
$\Large \psi^{(Z)}$ and inverts $\Large \bf a_Z$ into a \un
{pseudo Nambu-Goldstone boson (PNGB)}.

\clearpage\newpage

\subsection*{\un{The axion potential.}}

\vspace{1cm}

Then the field $\Large \bf \varphi_Z$ becomes:
 $$ \Large \bf
      \varphi_Z = \exp(ia_Z/v_Z)(v_Z + \sigma(x))\approx
           v_Z + \sigma(x) +ia_Z(x),
$$
with the field $\Large \bf \sigma $ is an \un{inflaton}.

Our axion $\Large \bf a_Z(x)$ is just the famous PQ-axion with a
mass squared
$$\Large \bf m_a^2\sim \Lambda_Z^3/v_Z\sim 10^{-30}\,\, Gev^2,$$
and its potential, given by PQ model, has (for small $\Large \bf
a_Z$) the expression of the following type:
$$\Large \bf
    V_{axion}\approx  \frac{\lambda}{4}(\varphi^+_Z\varphi_Z - v_Z^2)^2 + K
    |\varphi|\cos(a_Z/v_z),
$$
where $\Large \bf K$ is a positive constant: $\Large \bf K>0$.\\[3mm]
It is well-known that this potential exhibits two degenerate
minima at $\Large \bf <a_Z> = 0$ and $\Large \bf <a_Z> = 2\pi v_Z$
with the potential barrier existing between them.\\[3mm]
The minimum of the above-mentioned potential at $\Large \bf <a_Z>
= 0$ corresponds to the 'true' vacuum, while the minimum at
\\ $\Large \bf <a_Z> = 2\pi v_Z$ is called the 'false' vacuum.\\[3mm]
Such properties of the present axion leads to the 'quintessense'
model of our expanding universe and the axion $\Large \bf a_Z$
could be called an \un {acceleron}.

\clearpage\newpage

\subsection{\un{A new cosmological scale $\large \bf \Lambda_Z\approx
3\times 10^{-3}$ eV.}}

A new gauge group $\Large \bf SU(2)'_Z$ introduces a new dynamical
scale $\Large \bf \Lambda_Z\sim 10^{-3}$ eV, which is consistent
with present
measurements of cosmological constant:\\[3mm]
{\Large \bf \it A.G.~Riess et al.,} Astron.J. {\bf 116}, 1009
(1998); ArXiv: astro-ph/9805201.\\
{\Large \bf \it S.J.~Perlmutter et al.,} Nature {\bf 39}, 51
(1998); Astrophys.J. {\bf 517}, 565 (1999).\\
{\Large \bf \it C.~Bennett et al.,} ArXiv: astro-ph/0302207.\\
{\Large \bf \it D.~Spergel et al.,} ArXiv: astro-ph/0302209.\\
{\Large \bf \it P.~Astier et al.,} ArXiv: astro-ph/0510447.\\
{\Large \bf \it D.~Spergel et al.,} ArXiv: astro-ph/0603449.\\[3mm]
A total vacuum energy density of our universe (named cosmological
constant) is equal to the following value:$$ \Large \bf
\rho_{vac}\approx (3\times 10^{-3}\,\,\rm{eV})^4.$$

A new asymptotically free gauge group $\Large \bf SU(2)'_Z$ gives
the running of its inverse fine structure constant $\Large \bf
{(\alpha')}_{2Z}^{-1}(\mu)$, which grows from the extremely low
energy scale $\Large \bf \Lambda_Z\sim 10^{-3}$ eV up to the
supersymmetric scale $\Large \bf M_{SUSY}$ and then continues to
run (in our model -- does not change, see Figs.~1,2)\, up to the
superGUT scale
$\Large \bf M'_{SGUT} = M_{E6}\sim 10^{18}$ GeV.\\[3mm]
Near the scale $\Large \bf \Lambda_Z\sim 10^{-3}$ eV the coupling
constant of the gauge group $\Large \bf SU(2)'_Z$ infinitely
grows. At this scale we have a minimum of the effective potential
(the first vacuum in the mirror world).

Now it is worth the reader's attention to observe that in the
mirror world we have three scales (presumably corresponding to the
three vacua of the universe): $$\Large \bf \Lambda_1 =
\Lambda_Z\sim 10^{-12}\,\,\rm {\Large \bf GeV}, \quad \Large \bf
\Lambda_2 = \Lambda_{EW}\sim 10^3 \,\,\rm{\Large \bf GeV,}$$ $$
\Large \bf \Lambda_3 = \Lambda_{SGUT}\sim 10^{18}\,\, \rm {\Large
\bf GeV}.$$  They obey the following interesting relation:
$$\Large \bf \Lambda_1\cdot \Lambda_3 \approx \Lambda_2^2.$$

\clearpage\newpage

\section{\un{The gauge group $\Large \bf SU(2)'_Z$ and the 'quintessence'
model}} {\Large \bf \un {of our universe.}}

\vspace{5mm}

Recent models of the Dark Energy (DE) and Dark Matter (DM) are
based on measurements in contemporary cosmology.\\[3mm]
 Supernovae observations at redshifts
($\Large \bf 1.25 \le z \le 1.7$) by the Supernovae Legacy Survey
(SNLS), cosmic microwave background (CMB), cluster data and baryon
acoustic oscillations by the Sloan Digital Sky Survey (SDSS) fit
the equation of state for DE: $$\Large \bf  w = p/\rho$$ with a
constant $\Large \bf  w$:
 $$\Large \bf  w = -1.023 \pm 0.090 \pm 0.054.$$
which is given by \\[5mm]
{\Large \bf \it P.~Astier et al.,} ArXiv: astro-ph/0510447.\\[3mm]
 The value $\Large \bf w=-1$ is consistent with the present quintessence model of
accelerating universe  dominated by a tiny cosmological constant
and Cold Dark Matter (CDM):\\[3mm]
{\Large \bf \it P.J.E.~Peebles and A.~Vilenkin,} Phys.Rev. D {\bf
59},
063505 (1999),\\
 {\Large \bf \it C.~Wetterich,} Nucl.Phys. B {\bf
302}, 668 (1998),\\
{\Large \bf \it L.J.~Hall, Y.~Nomura, S.J.~Oliver,} Phys.Rev.Lett.
{\bf 95}, 141302 (2005); ArXiv: astro-ph/0503706.\\[5mm]
Here we present the quintessence scenario, which was developed in
connection with the existence of a new gauge group $\Large \bf
SU(2)'_Z$.\\[3mm]
In our model $\Large \bf a_Z$ plays the role of the 'acceleron',
and a scalar boson $\Large \bf \sigma_Z$, partner of the
acceleron, plays the role of the 'inflaton' in the low scale
inflationary scenario.

\clearpage\newpage

\subsection{\un{Dark energy and cosmological constant.}}

For the ratios of densities $\Large \bf \Omega_X =\rho_X/\rho_c$,
cosmological measurements gave: $$\Large \bf \Omega_B\sim 4\%$$
for baryons (visible and dark), $$\Large \bf \Omega_{DM}\sim
23\%$$ for non-baryonic DM, and $$\Large \bf \Omega_{DE}\sim
73\%$$ for the mysterious DE, which is responsible for the
acceleration of our
universe.\\[3mm]
We have considered that a cosmological constant (CC) is given by
the value  $$\Large \bf CC = \rho_{vac}\approx (3\times
10^{-3}\,\,\rm{eV})^4.$$ The main assumption is the following
idea:\\[3mm]
the universe is trapped in the false vacuum with $CC\neq 0$, but
at the end it must decay into the true vacuum with vanishing
CC.\\[3mm]
The true Electroweak vacuum would have its vacuum energy density
$$\Large \bf CC = \rho_{vac} = 0.$$ Such a scenario exists in the
model with Multiple Point Principle (MPP):\\[3mm]
 {\Large \bf \it
D.L.~Bennett, H.B.~Nielsen,} Int.J.Mod.Phys. A {\bf 9}, 5155
(1994); ibid., A {14}, 3313 (1999).\\See review:
 {\Large \bf \it C.R.~Das, L.V.~Laperashvili,}
Int.J.Mod.Phys. A {\bf 20}, 5911 (2005).\\[3mm]
A non-zero (nevertheless tiny) CC would be associated only with a
false vacuum.\\[3mm]
{\Large \bf Why CC is zero in a true vacuum?}\\[3mm]

\clearpage\newpage

\subsection*{\un{Dark energy and cosmological constant.}}

People try to give a solution of this non-trivial problem:\\[3mm]
{\Large \bf \it C.D.~Froggatt, L.V.~Laperashvili, R.B.~Nevzorov,
H.B.~Nielsen,} in: Proceedings of 7th Workshop on 'What Comes
Beyond the Standard Model', Bled, Slovenia, 19-30 Jul 2004 (DMFA,
Zaloznistvo, Ljubljana, 2003), p.17; ArXiv: hep-ph/0411273.\\[3mm]
{\Large \bf \it L.~Mersini,} ArXiv: gr-qc/0609006.\\[3mm]
The axion potential $\Large \bf V(a_Z)$ determines the origin of
DE:\\[3mm]
when the temperature of the universe $\Large \bf T$ is high:
$\Large \bf T >> \Lambda_Z$, then the axion potential is flat
because the effects of the $\Large \bf SU(2)'_Z$ instantons are
negligible for such temperatures.\\[3mm]
When the temperature begins to decrease, the universe gets trapped
in the false vacuum.\\[3mm]
At $\Large \bf T\sim \Lambda_Z$ the true vacuum at $\Large \bf
<a_Z> = 0$ has zero energy density (cosmological constant), and
the barrier between vacua is higher.\\[3mm]
The difference in energy
density between the true and false vacua is now $\Large \bf
\Lambda_Z^4$. The universe is still trapped in the false vacuum
with $\Large \bf CC = \rho_{vac} = \Lambda_Z^4$.\\[3mm]
The first order phase transition to the true vacuum is provoked by
the bubble nucleation. In fact, the universe lives in the false
vacuum for a very long time.\\[3mm]
When the universe is trapped in the false vacuum at $$\Large \bf
<a_Z> = 2\pi v_Z,$$ the deceleration stops and acceleration begins
at $\Large \bf \ddot a_Z = 0$,\\[3mm]
then $\Large \bf {\dot a}^2_Z = 0$ and $\Large \bf
w(a_Z) = -1$. \\[3mm] The total energy density of the universe is
dominated by the energy density of the false vacuum, and our
universe undergoes an exponential expansion.\\[3mm]
The universe trends to get the true vacuum, which has zero CC.

\clearpage\newpage

\subsection{\un{Dark matter.}}

The existence of DM (non-luminous and non-absorbing matter) in the
universe is now well established.\\[3mm] Candidates for non-baryonic DM
must be particles, which are stable on cosmological time scales.
They must interact very weakly with electromagnetic radiation.
Also they must have the right relic density.\\[3mm] These candidates can
be black holes, axions, and weakly interacting massive particles
(WIMPs). In supersymmetric models WIMP candidates are the lightest
superparticles. The most known WIMP is the lightest neutralino.
WIMPs could be photino, higgsino, or bino.\\[3mm]
In our model fermions $\Large \bf \psi^{(Z)}_i$ of the gauge group
$\Large \bf SU(2)'_Z$ also could be considered as candidates for
HDM (hot dark matter), and their composites ("hadrons" of $\Large
\bf SU(2)_Z$) could play a role of the WIMPs in CDM.\\[3mm]
Investigating DM, it is possible to search and study various
signals such as: $\Large \bf \psi^{(Z)} + e \to \psi^{(Z)} + e$,
or $\Large \bf \psi^{(Z)} + N \to \psi^{(Z)} + N$, where $\Large
\bf e$ is an electron and
$\Large \bf N$ is a nucleon.\\[3mm]
The detection of mirror particles: mirror quarks, leptons, Higgs
bosons, etc., could be performed at future colliders such as LHC.\\[3mm]
Also the 'messenger' scalar boson $\phi^{(Z)}$ can be produced at
LHC, and then the decay: $\phi_i^{(Z)} \to \ov{\psi}^{(Z)}_i + l$,
where $\Large \bf l$ stands for the SM lepton, can be investigated
with $\Large \bf \psi^{(Z)}$ as missing energies.\\[3mm]
Leptogenesis and inflationary model also can be considered as
implications of our new physics. The full investigation is beyond
this paper.

\clearpage\newpage

\section{\un{Conclusions.}}

\begin{itemize}

\item[{\bf 1.}] In this talk we have discussed cosmological
implications of the parallel ordinary and mirror worlds with the
broken mirror parity MP.

\item[{\bf 2.}] We have considered the parameter characterizing
the breaking of MP, which is $\Large \bf \zeta = v'/v$, where
$\Large \bf v'$ and $\Large \bf v$ are the VEVs of the Higgs
bosons -- Electroweak scales -- in the M- and O-worlds,
respectively.

\item[{\bf 3.}] During our numerical calculations, we have used
the value $\Large \bf \zeta\approx 30$, in accordance with a
cosmological estimate obtained by Berezhiani, Dolgov and
Mohapatra.

\item[{\bf 4.}] We have assumed that at the very small distances
there exists $\Large \bf E_6$-unification predicted by Superstring
theory. We have chosen a theory, which leads to the asymptotically
free $\Large \bf E_6$ unification, what is not always fulfilled.

\item[{\bf 5.}] We have shown that, as a result of the
MP-breaking, the evolutions of fine structure constants in O- and
M-worlds are not identical, and the extensions of the Standard
Model in the ordinary and mirror worlds are quite different.

\item[{\bf 6.}] We have assumed that the $\Large \bf
E_6$-unification, being the same in the O- and M-worlds, restores
the broken mirror parity MP.

\item[{\bf 7.}] We have considered the following chain of symmetry
groups in the ordinary world:
$$\Large \bf  SU(3)_C\times
SU(2)_L\times U(1)_Y$$ $$\Large \bf \to
 SU(3)_C\times SU(2)_L \times SU(2)_R\times U(1)_X\times
 U(1)_Z$$
 $$\Large \bf \to SU(4)_C\times SU(2)_L \times SU(2)_R\times U(1)_Z\to
SO(10)\times U(1)_Z \to E_6.$$

\clearpage\newpage

\section*{\un{Conclusions.}}

\item[{\bf 8.}] We have shown that a simple logic leads to the
following chain in the mirror world:
$$\Large \bf SU(3)'_C\times SU(2)'_L\times SU(2)'_Z\times U(1)'_Y$$
$$\Large \bf \to SU(3)'_C\times SU(2)'_L\times SU(2)'_Z\times U(1)'_X\times U(1)'_Z$$
 $$\Large \bf \to
SU(4)'_C\times SU(2)'_L\times SU(2)'_Z\times U(1)'_Z $$ $$\Large
\bf \to SU(6)'\times SU(2)'_Z \to E'_6.$$

\item[{\bf 9.}] The comparison of both evolutions in the ordinary
and mirror worlds is given in Figs.~1--4, where we have presented
the running of all fine structure constants. Here the SM (SM') is
extended by MSSM (MSSM'), and we see different evolutions.
Figs.~1,2 correspond to the SUSY breaking scales$$\Large \bf M_{SUSY}=10\,\,TeV,\quad \Large \bf
M'_{SUSY}=300\,\,TeV,$$ while Figs.~3,4 are presented for
$$\Large \bf M_{SUSY}=1\,\, TeV, \quad \Large \bf
M'_{SUSY}=10\,\,TeV,$$ according to the MP-breaking parameter
$\Large \bf \zeta\approx $30 and 10. We have considered the value
of seesaw scale in the O-world
$$\Large \bf M_R\sim 10^{14}\,\, GeV,$$ and in the M-world:$$\Large \bf M'_R\sim 10^{17}\,\, GeV.$$

\item[{\bf 10.}] It was shown that the (super)grand unification
$\Large \bf E'_6$ in the mirror world is based on the group
$$\Large \bf E'_6\supset SU(6)'\times SU(2)'_Z.$$

\item[{\bf 11.}] The presence of a new gauge group $\Large \bf
SU(2)'_Z$ in the M-world gives significant consequences for
cosmology: it explains the 'quintessence' model of our
accelerating universe.

\clearpage\newpage

\section*{\un{Conclusions.}}

\item[{\bf 12.}] We have presented in Figs.~1--4 the running of
the $\Large \bf SU(2)'_Z$ gauge coupling by the evolution $\Large
\bf {\alpha'}_{2Z}^{-1}(\mu)$, which takes its initial value at
the superGUT scale $$\Large \bf M_{SGUT}\sim 10^{18}\,\,GeV$$ and
then runs down to very low energies, giving an extremely strong
coupling constant at the scale $\Large \bf \Lambda_Z\sim 10^{-3}$
eV.

\item[{\bf 13.}] We have discussed  a 'quintessence' model of our
universe: at the scale $\Large \bf \Lambda_Z \sim 10^{-3}$ eV
instantons of the gauge group $\Large \bf SU(2)'_Z$ induce a
potential for an axion-like scalar boson $\Large \bf a_Z$, which
can be called "acceleron". The acceleron gives the value $\Large
\bf w=-1$ and leads to the acceleration of our universe.

\item[{\bf 14.}] It was shown that the existence of the scale
$\Large \bf \Lambda_Z\sim 10^{-3}$ eV explains the value of
cosmological constant: $$\Large \bf CC\approx (3\times
10^{-3}\,\,\rm{\Large \bf  eV})^4,$$ which is given by
astrophysical measurements. Also recent measurements in cosmology
fit the equation of state for DE: $\Large \bf  w = p/\rho$ with a
constant $\Large \bf w\approx -1$.

\item[{\bf 15.}] Following P.Q.~Hung, we have assumed that at
present time our universe exists in the 'false' vacuum given by
the axion potential. The universe will live there for a long time
and its CC (measured in cosmology) is tiny, but nonzero. However,
at the end the universe will jump into the 'true' vacuum and will
get a zero CC. But this problem is not trivially solved, and at
present time we have only a hypothesis.

\item[{\bf 16.}] It was observed that the mirror world has three
scales:
$$\Large \bf \Lambda_1 = \Lambda_Z\sim 10^{-12}\,\,
\rm {\Large \bf GeV}, \quad \Large \bf \Lambda_2 =
\Lambda_{EW}\sim 10^3 \,\,\rm{\Large \bf GeV},$$ $$\Large \bf
\Lambda_3 = \Lambda_{SGUT}\sim 10^{18}\,\, \rm {\Large \bf GeV}.$$
They obey the following interesting relation:
$$\Large \bf \Lambda_1\cdot \Lambda_3 \approx \Lambda_2^2.$$

\clearpage\newpage

\section*{\un{Conclusions.}}

\item[{\bf 17.}] In our model of the universe with broken mirror
parity we have obtained the following particle content of the
group $\Large \bf SU(2)'_Z$:

$\bullet$  two doublets of fermions $\Large \bf \psi^{(Z)}_i$
($\Large \bf i = 1,2$),\\
         or a triplet of fermions $\Large \bf \psi^{(Z)}_f$  ($\Large \bf f = 1,2,3$);

$\bullet$  two doublets of scalar fields $\Large \bf \phi^{(Z)}_i$
($\Large \bf i = 1,2$).

\item[{\bf 18.}] Also, as H.~Goldberg and P.Q.~Hung, we have
considered an existence of a complex singlet scalar field $\Large
\bf \varphi_Z$, which produces "acceleron" $\Large \bf a_Z$ and
"inflaton" $\Large \bf \sigma_Z$ and gives a 'quintessence' model
of our universe with the low scale inflationary scenario.

\item[{\bf 19.}] Unfortunately, we cannot predict exactly the
scales $\Large \bf M_{SUSY}$ and $\Large \bf M_R$ presented in our
Figs.~1--4. The numerical description of the model depends on
these scales. Nevertheless, we hope that a qualitative scenario
for the evolution of our universe, developed in this paper, is
valid.

\item[{\bf 20.}] We have discussed a possibility to consider the
fermions $\Large \bf \psi^{(Z)}_i$ of the group $\Large \bf
SU(2)'_Z$ as candidates for HDM and composites ("hadrons" of
$\Large \bf SU(2)'_Z$) as WIMPs in CDM. Searching DM, it is
possible to observe and study various signals of these particles.

\item[{\bf 21.}] Finally, it is necessary to emphasize that this
investigation opens the possibility to fix a grand unification
group ($\Large \bf E_6$ ?) from cosmology.

\end{itemize}

\normalsize

\section{Acknowledgements}

C.R.D. deeply thanks Prof.~R.N.Mohapatra for useful advices.

This work was supported by the Russian Foundation for Basic
Research (RFBR), project No 05–-02–-17642.

\clearpage\newpage \bfi \centering
\includegraphics[height=80mm,keepaspectratio=true,angle=0]{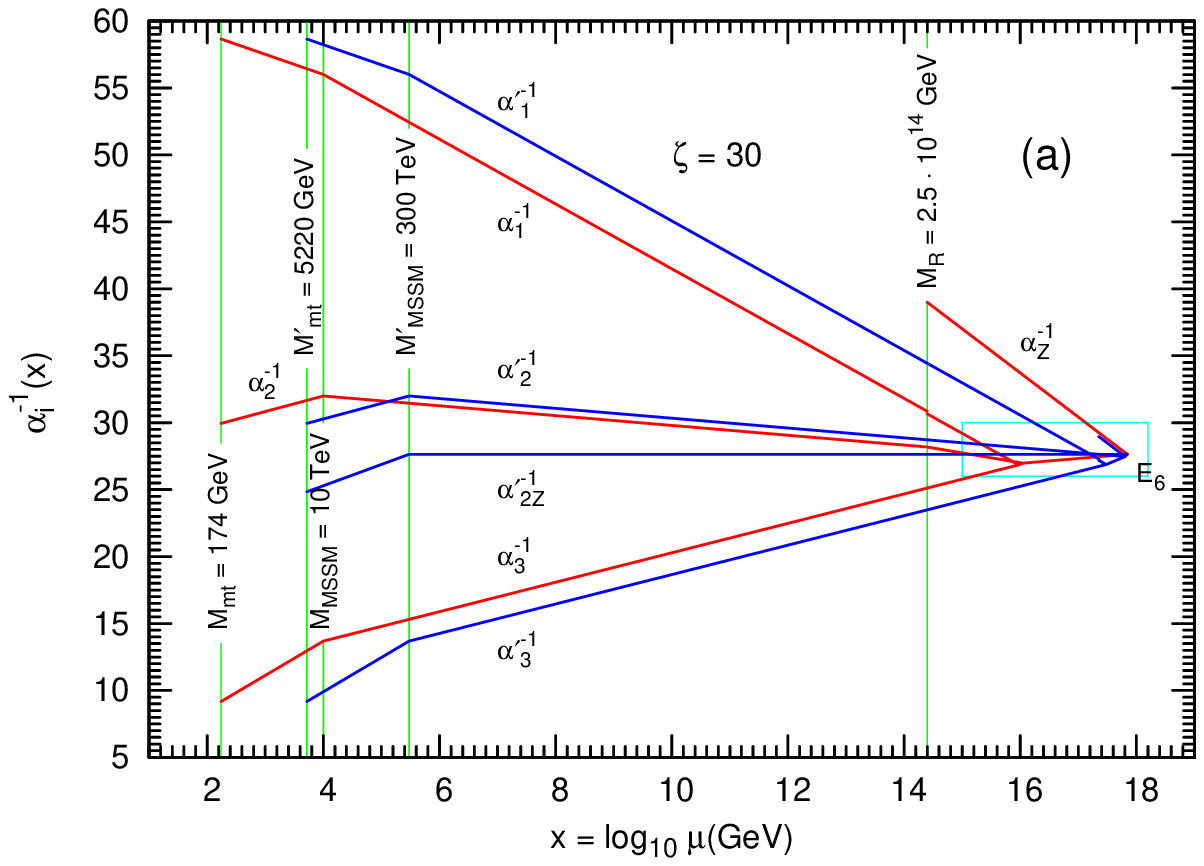} \caption
{The running of the inverse coupling constants $\alpha_i^{-1}(x)$
in both ordinary and mirror worlds with broken mirror parity from
the Standard Model up to the $E_6$ unification for SUSY breaking
scales $M_{SUSY}= 10$ TeV, $M'_{SUSY}= 300$ TeV; $\zeta =30;$ and
seesaw scales $M_R=2.5\cdot 10^{14}$ GeV, $M'_R=2.25\cdot 10^{17}$
GeV. This case gives: $M_{SGUT}\approx 7\cdot 10^{17}$ GeV and
$\alpha_{SGUT}^{-1}\approx 27.64$.} \efi

\bfi \centering
\includegraphics[height=80mm,keepaspectratio=true,angle=0]{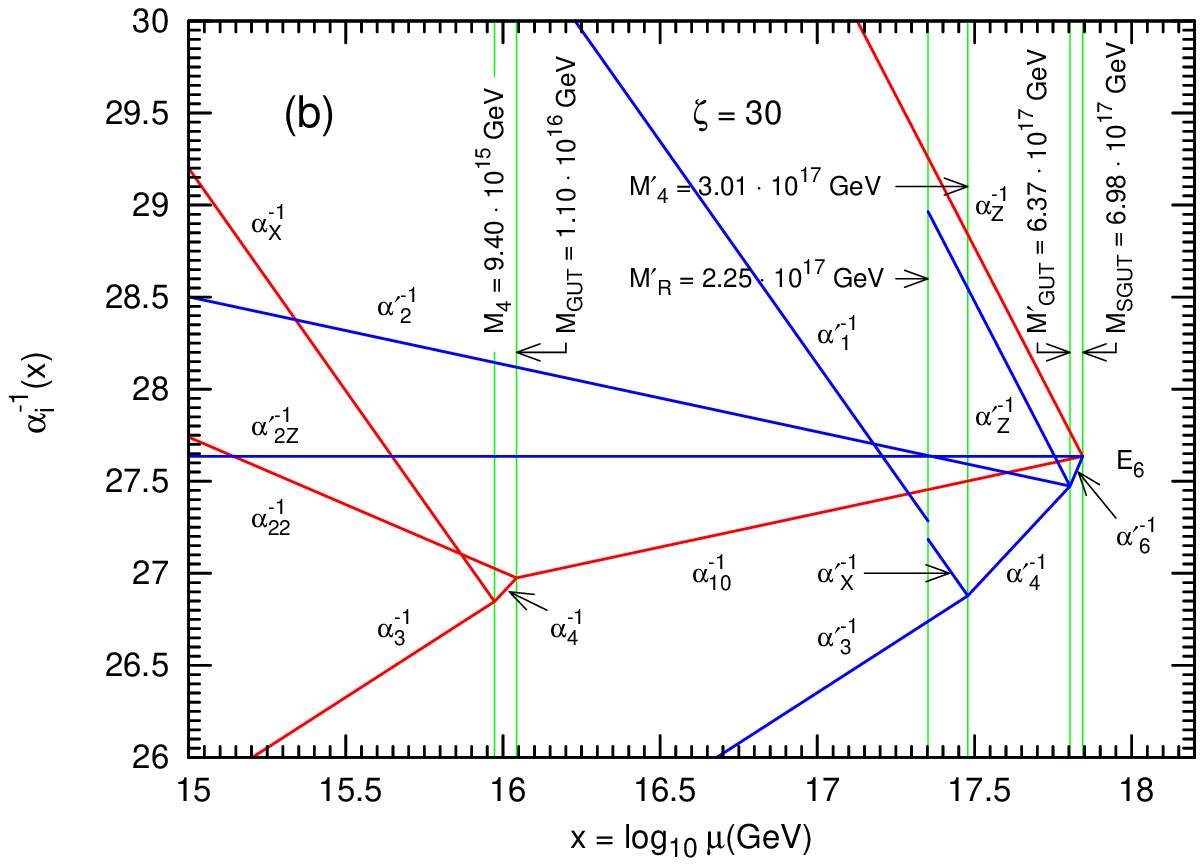}
\caption { This figure presents the same running of the inverse
coupling constants $\alpha_i^{-1}(x)$ in both ordinary and mirror
worlds from the scale $10^{15}$ GeV up  to the $E_6$ unification
for SUSY breaking scales $M_{SUSY}= 10$ TeV, $M'_{SUSY}= 300$ TeV;
$\zeta =30$; and seesaw scales $M_R=2.5\cdot 10^{14}$ GeV,
$M'_R=2,25\cdot 10^{17}$ GeV; $M_{SGUT}\approx 7\cdot 10^{17}$ GeV
and $\alpha_{SGUT}^{-1}\approx 27.64$.} \efi

\clearpage\newpage \bfi \centering
\includegraphics[height=80mm,keepaspectratio=true,angle=0]{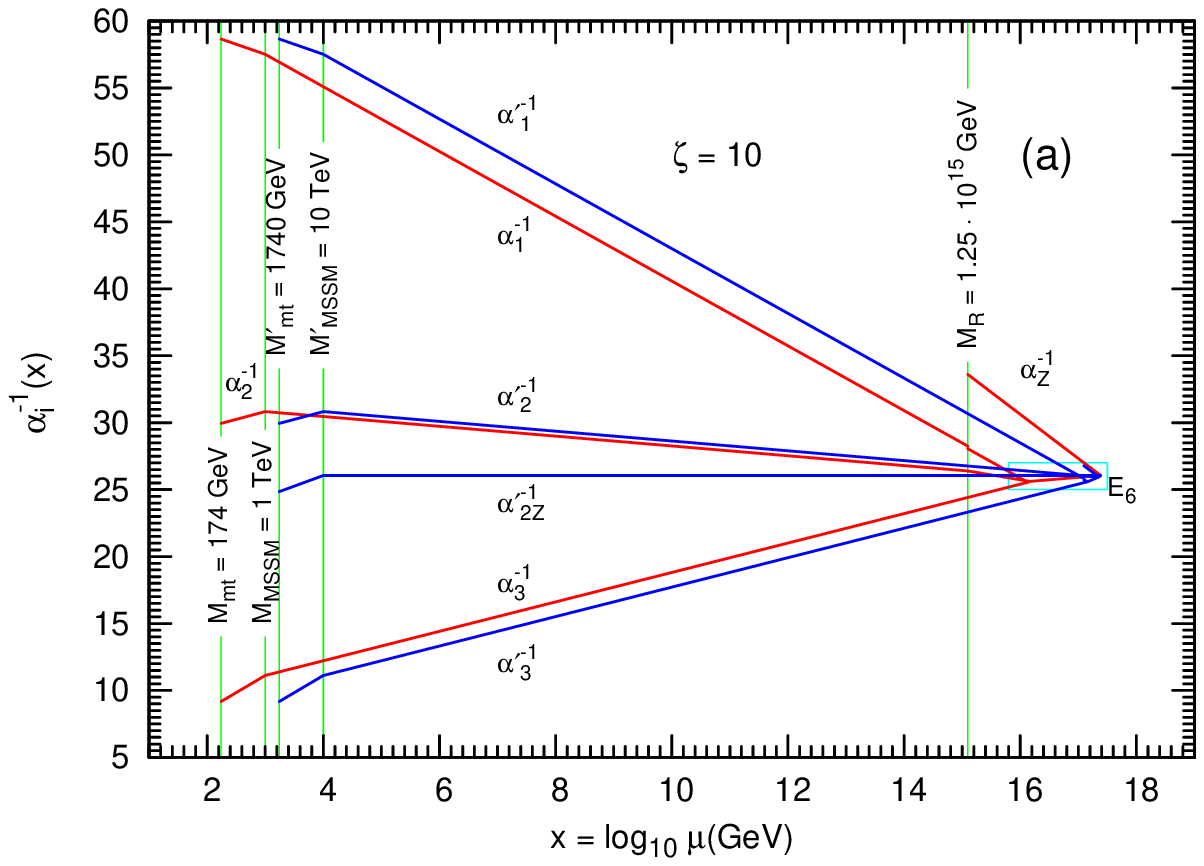}
\caption { The running of the inverse coupling constants
$\alpha_i^{-1}(x)$ in both ordinary and mirror worlds with broken
mirror parity from the Standard Model up to the $E_6$ unification
for SUSY breaking scales $M_{SUSY}= 1$ TeV, $M'_{SUSY}= 10$ Te;
$\zeta =10$; and seesaw scales $M_R=1.25\cdot 10^{15}$ GeV,
$M'_R=1.44\cdot 10^{17}$ GeV. This case gives: $M_{SGUT}\approx
2.4\cdot 10^{17}$ GeV and $\alpha_{SGUT}^{-1}\approx 26.06$.} \efi

\bfi \centering
\includegraphics[height=80mm,keepaspectratio=true,angle=0]{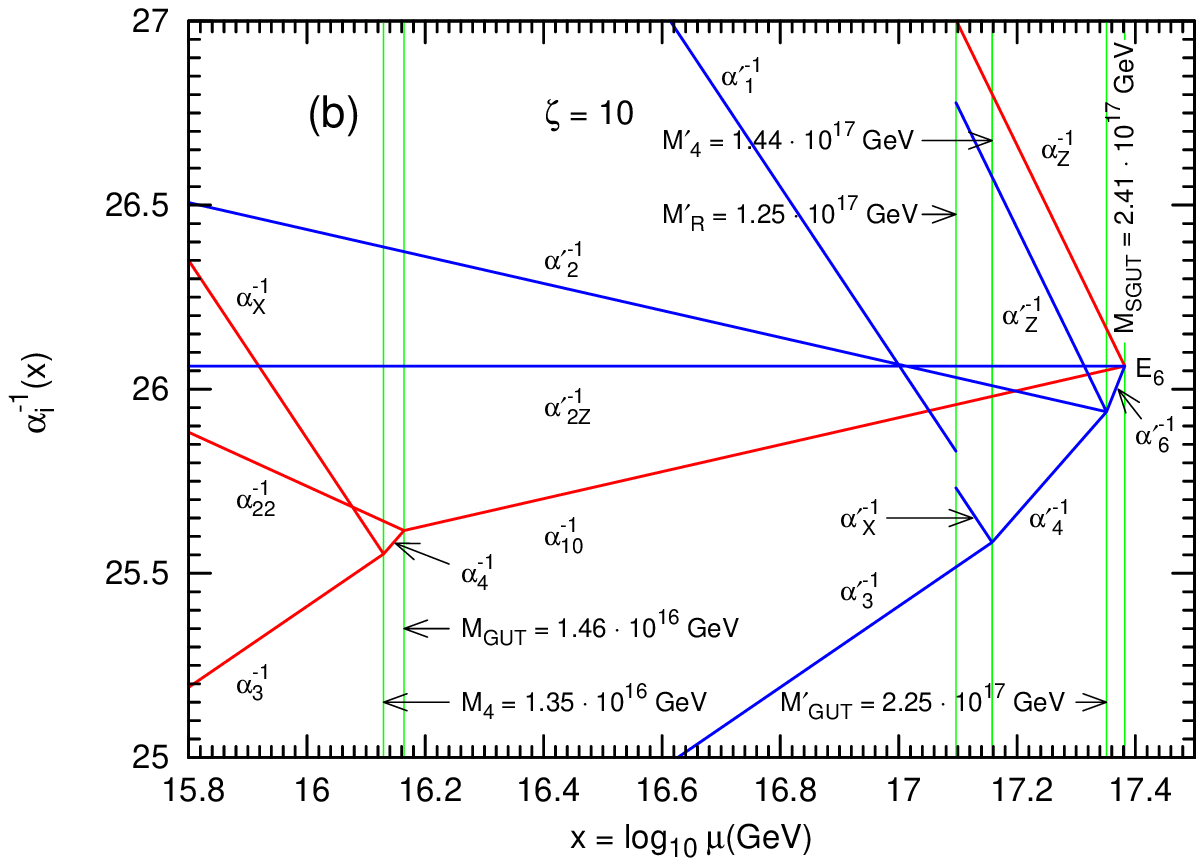}
\caption { This figure presents the same running of the inverse
coupling constants $\alpha_i^{-1}(x)$ in both ordinary and mirror
worlds with broken mirror parity from the scale $10^{16}$ GeV up
to the $E_6$ unification for SUSY breaking scales $M_{SUSY}= 1$
TeV, $M'_{SUSY}= 10$ TeV; $\zeta =10$; and seesaw scales
$M_R=1.25\cdot 10^{15}$ GeV, $M'_R=1.44\cdot 10^{17}$ GeV;
$M_{SGUT}\approx 2.4\cdot 10^{17}$ GeV and
$\alpha_{SGUT}^{-1}\approx 26.06$.} \efi

\end{document}